\documentclass[sigplan,nonacm]{acmart}
\usepackage[utf8]{inputenc}
\usepackage{href-ul}
\usepackage{natbib}
\usepackage{graphicx}
\usepackage{amsmath}
\usepackage[noframes]{ffcode}
\usepackage{to-be-determined}
\usepackage{etoolbox}
\usepackage{paralist}
\usepackage{tabularx}
\usepackage{booktabs}
\usepackage[capitalize]{cleveref}

\title{The Impact of Mutability on Cyclomatic Complexity in Java}

\author{Marat Bagaev}
\orcid{0009-0002-5292-9404}
\email{mtbagaev@edu.hse.ru}
\affiliation{\institution{HSE}\city{Moscow}\country{Russia}}

\author{Alisa Khabibrakhmanova}
\orcid{0009-0000-6250-8385}
\email{aakhabibrakhmanova@hse.ru}
\affiliation{\institution{HSE}\city{Moscow}\country{Russia}}

\author{Georgy Sabaev}
\orcid{0009-0001-3022-1686}
\email{gasabaev@hse.ru}
\affiliation{\institution{HSE}\city{Moscow}\country{Russia}}

\author{Yegor Bugayenko}
\orcid{0000-0001-6370-0678}
\email{yegor256@gmail.com}
\affiliation{\institution{Huawei}\city{Moscow}\country{Russia}}

\newcommand{\mtc}[1]{\text{\scshape#1}}

\begin{document}

\begin{abstract}
In Java, some object attributes are mutable, while others are immutable (with the "final" modifier attached to them). Objects that have at least one mutable attribute may be referred to as "mutable" objects. We suspect that mutable objects have higher McCabe's Cyclomatic Complexity (CC) than immutable ones. To validate this intuition, we analysed 862,446 Java files from 1,000 open-GitHub repositories. Our results demonstrated that immutable objects are almost three times less complex than mutable ones. It can be therefore assumed that using more immutable classes could reduce the overall complexity and maintainability of the code base.
\end{abstract}

\maketitle

\section{Introduction}

In the field of software development, mutable data is common way to represent the internal state of a program. Many algorithms and data structures, such as in-place sorting algorithms and linked lists, are made with mutable data in mind. Support for mutability is provided by an overwhelming majority of popular programming languages, including Python, Java, C\#, C++, Kotlin and many others, and is especially common in languages and programs that embrace the paradigm of Object Oriented Programming (OOP).

However, some experts consider mutability a harmful practice and recommend avoiding it as much as possible~\citep{west2004object,goetz2006java,bloch2017,eo1}. Some common arguments against mutable data include:
\begin{inparaenum}[a)]
    \item difficulties caused by concurrency, such as race conditions
    and
    \item a possibility of an object entering an invalid state during code execution.
\end{inparaenum}
This implies that, while the concept of mutability is by itself very familiar to most programmers, working with it may often introduce additional complexity of extra sanity checks that could significantly bloat the codebase.

With this in mind, it is important to assess and quantify the impact of mutability on code complexity, to see if minimising or outright rejecting it could result in cleaner and more concise code. We suspect that the lack of direct evidence connecting mutable data and code complexity contributes to slow adoption of immutable data in code. Without such solid evidence, it is hard to justify learning and switching to a different approach to data storage and rewriting existing code to match it.

In this research, we aim to analyze open source Java repositories in GitHub and answer the following research questions:
\begin{description}
    \item[RQ1] What is the average ratio between mutable and immutable objects in Java projects?
    \item[RQ2] What is the correlation between the percentage of mutable attributes in a Java class and the \mtc{CC} of the class?
    \item[RQ3] Do mutable objects tend to have higher \mtc{CC} than immutable ones?
\end{description}
Our thesis statement is that immutability favourably affects simplicity of a Java class. If this hypothesis holds true, it can help software engineers make more informed decisions about their system architecture by minimising mutability within the codebase.

This article is structured as follows: 
\cref{sec:background} explains the most important concepts used in the research, 
\cref{sec:related} presents a review of works related to the research, 
\cref{sec:method} outlines practical steps taken during the study, 
\cref{sec:results} describes obtained results, 
\cref{sec:discussion} provides interpretations of our findings and explores limitations,
\cref{sec:limitations} highlights limitations of our study and threats to its validity, and 
\cref{sec:conclusion} offers a summary of the paper.

\section{Background}\label{sec:background}

Cyclomatic Complexity (\mtc{CC}), introduced by \citet{mccabe1976}, is a metric aimed to assess and quantify the number of possible paths through a function, and increases with any branching in the logical flow of the program. For a function, \mtc{CC} is calculated using the following formula:
\begin{equation*}
C = E - N + 2,
\end{equation*}
where $N$ is the number of nodes of the function's control flow graph (including the entry/exit nodes) and $E$ is the number of edges of the function's control flow graph (nodes are connected with an edge if the program flow can reach one from the other directly). 

The \mtc{CC} formula can be extended to assess the complexity of an entire class across all of its methods, by making a summary of \mtc{CC}'s for each individual method of the class. This is the metric we use in order to evaluate the complexity of Java classes in the research.

Informally, in object-oriented programming, \emph{mutable} objects are the objects that contain (non-static) mutable attributes, which are possible to modify after the object is instantiated. The objects that are not mutable are called \emph{immutable}.

In the Java language specification, immutable objects are not formally defined~\citep{immutable_java_spec}. Instead, the documentation provides the following guidelines to make an object immutable:
\begin{itemize}
\item Don't provide ``setter'' methods---methods that modify fields or objects referred to by fields;
\item Make all fields final and private.
\item Don't allow subclasses to override methods. The simplest way to do this is to declare the class as final. A more sophisticated approach is to make the constructor private and construct instances in factory methods.
\item If the instance fields include references to mutable objects, don't allow those objects to be changed:
\item Don't provide methods that modify the mutable objects.
\item Don't share references to mutable objects. Never store references to external, mutable objects passed to the constructor; if necessary, create copies, and store references to the copies. Similarly, create copies of your internal mutable objects when necessary to avoid returning the originals in your methods.
\end{itemize}
A simpler way to roughly estimate whether a given Java class is immutable or not is to check whether it has any non-static non-final attributes

\section{Related Work}\label{sec:related}

Some experts suggest using immutable classes in programs instead of mutable ones whenever possible~\citep{goetz2006java,bugayenko2014blog0609}. \citeauthor{bloch2017}, one of the creators of Java, recommends minimizing mutability of objects~\citep{bloch2017}: ``Classes should be immutable unless there’s a very good reason to make them mutable.'' It is a known belief that mutability in objects can cause severe security and code usability problems~\citep{weber2017empirical}. 

There were studies analyzing correlation between CC and some other metrics. For example, \citet{meine2007correlations} and \citet{graylin2009cyclomatic} demonstrated the presence of a strong linear correlation between CC and Lines of Code (LoC); \citet{muslija2018correlation} showed that a correlation between the effort needed to test a program and its complexity is moderate; \citet{seront2005relationship} observed no significant correlation between the depth of inheritance of a class and its weighted method complexity; \citet{abd2018} showed that the probability of the occurrence or emergence of new errors positively correlates with the CC; \citet{shin2008complexity} found a weak correlation between code vulnerability and its complexity; \citet{hindle2008reading} discovered a correlation between CC and the depth of indentation; \citet{mamun2017correlations} identified a correlation between complexity and the quality of source code documentation.

There was a study of \citet{bugayenko2020impact} that demonstrated that immutable classes are more cohesive. They suggested that, in object-oriented design, immutability positively contributes to the quality of code.

Problems of usage of immutable objects pushed some researchers to study immutability and find ways to simplify the usage of immutable classes. An example of such a method is a type annotation system for Java called Glacier \citep{coblenz2017glacier}, which provides, among other features, the ability to explicitly mark a class as |@Immutable| and then check in runtime whether objects are truly immutable.

To this date, there is no work considering the correlation between mutability of Java classes and their complexity.

\section{Method}\label{sec:method}

To answer our research questions, we collected data about existing code repositories and then analysed it to extract important metrics and observe the way these metrics and \mtc{CC} depend on one another.

As the base of our experimental data, we used the ``22-10-2023'' version of an existing CaM dataset~\citep{cam2024}. The dataset was generated using publicly available code, which we took as a base. The original script, when it is launched, downloads 1000 Java repositories with 1k-10k stars and with at least 200~KB of code and analyses the abstract syntax tree (AST) of each individual Java class, looking for several metrics that are specified in the code.

However, not all the required metrics were provided; namely, the number of mutable attributes was not in the dataset. Hence, we added the following metrics to the script and reran the data collection on our own:
\begin{itemize}
\item \mtc{NoFA} --- number of final (immutable) attributes;
\item \mtc{NoSFA} --- number of final static attributes;
\item \mtc{Attrs} --- number of non-static attributes;
\item \mtc{SAttrs} --- number of static attributes;
\item \mtc{FARatio} --- final attribute ratio, which is equal to $\mtc{NoFA} / (\mtc{Attrs} + \mtc{SAttrs})$;
\item \mtc{FARatioNS} --- final non-static attribute ratio, which is equal to $(\mtc{NoFA} - \mtc{NoSFA} ) / \mtc{Attrs}$.
\end{itemize}
The rest of the metrics defined earlier were then derived from these, and all the data was composed into a single dataset. 

After obtaining the dataset, we plotted \mtc{CC} against the ratio of \mtc{NoFA} to total number of attributes, with and without static attributes being taken into account. We looked for possible correlation on a scatter plot and checked if suspected correlations existed using Karl Pearson’s coefficient of correlation. We also calculated the mean and median \mtc{CC} of mutable and immutable objects (also with or without static attributes included) and compared them to see if any significant differences could be seen.

\section{Results}\label{sec:results}

We calculated the correlation between our metrics and got the following results (\cref{tab:correlations}).

\begin{table}
\caption{Low correlation between CC and either of the metrics that represent the extent of immutability.}
\label{tab:correlations}
\begin{tabularx}{\linewidth}{l>{\ttfamily}r>{\ttfamily}r>{\ttfamily}r}
\toprule
&{\rmfamily\mtc{CC}}  &	{\rmfamily \mtc{FARatio}} &{\rmfamily \mtc{FARatioNS}} \\
\midrule
\mtc{CC}	&1.0000		&0.0003	&-0.0204\\
\mtc{FARatio}	&0.0003		&1.0000	&0.9198\\
\mtc{FARatioNS} &-0.0204		 &0.9198 &1.0000\\
\bottomrule
\end{tabularx}
\end{table}

\Cref{tab:aggregated} shows the aggregated statistics of the dataset we collected.

\begin{table}
\caption{Purely mutable classes have significantly lower CC despite the lack of correlation with the extent of mutability.}
\label{tab:aggregated}
\begin{tabularx}{\linewidth}{X>{\ttfamily}r}
\toprule
Statistic & {\rmfamily Value} \\
\midrule
Number of immutable classes & 493,382 \\
Number of mutable classes & 369,064 \\
Average \mtc{CC} of immutable classes & 6.5 \\ 
Average \mtc{CC} of mutable classes & 19.8 \\
Median \mtc{CC} of immutable classes & 2.0 \\
Median \mtc{CC} of mutable classes & 5.0 \\
\bottomrule
\end{tabularx}
\end{table}

The scripts that we used are archived in Zenodo\footnote{\url{https://zenodo.org/records/10968240}}.

\section{Discussion}\label{sec:discussion}

One of the unresolved problems that we encountered during the data collection was accurately determining whether a class is mutable. The immutability criterion used in this paper is not a proper guarantee of immutability for many reasons, including, but not limited to, inheritance, reflection and the fact that a final attribute may be an object with exposed methods that modify their internal data. These difficulties are not unique to our research and are a general problem within the Java community, and tools such as Glacier~\citep{coblenz2017glacier} exist to mitigate it.

The flaw in our method comes from the fact that the dataset operates on every Java file individually, and has no reference for the other classes used within it. As such, there are several cases in which a class would be incorrectly classified as immutable.

\Cref{fig:false-immutable} shows an example of a class that would be counted as ``immutable'' despite mutable data being stored within it. Despite all attributes of the |FalseImmutable| class being flagged with the |final| keyword, the internal |List| has exposed methods (e.g. |List.add|), which can be used to modify its internal data.

\begin{figure}
\begin{ffcode}
class FalseImmutable {
  // Can be modified despite being final
  public final List<Integer> data; 
  private final int value;
}
var obj = new FalseImmutable();
obj.data.add(1);  // final attribute changed
\end{ffcode}
\caption{A final field can still conain mutable data.}
\label{fig:false-immutable}
\end{figure}

\Cref{fig:mutable-class} shows yet another example, where the problem arises from inheritance: the |FalseImmutable| class does not define any mutable fields in its definition, but extends the |MutableClass| class and therefore inherits a mutable attribute from it.

\begin{figure}
\begin{ffcode}
class MutableClass {
  public int mutableAttribute;
}
// will be incorrectly flagged as immutable
class FalseImmutable extends MutableClass {
  private final int immutableData;
}
var obj = new FalseImmutable();
// "immutable" object gets modified:
obj.mutableAttribute = 10;
\end{ffcode}
\caption{A final field can be inherited from the base class.}
\label{fig:mutable-class}
\end{figure}

\Cref{fig:immutable-class} shows a similar case for reflection, which works even for most well-formed immutable classes. The |ImmutableClass| class does not define any mutable fields in its definition; however, with the help of reflection, it is possible to modify its |final| attribute at runtime.

\begin{figure}
\begin{ffcode}
class ImmutableClass {
  public final int x;
}
var obj = new ImmutableClass();
Field fieldX = obj.getClass().getField("x");
fieldX.setAccessible(true);
fieldX.set(obj, 2);
\end{ffcode}
\caption{Even a properly immutable field can be modified through reflection.}
\label{fig:immutable-class}
\end{figure}

As evident from the data, purely immutable classes are significantly more common in the selected repositories, and their mean and median \mtc{CC} is drastically lower. However, when considering the ratio of immutable attributes, no significant correlation could be found.
We think that these results imply that the quality of a class being mutable at all is the important part for assessing complexity, and the degree of said mutability does not matter.

It can be therefore assumed that using more immutable classes could reduce the overall \mtc{CC} of the code base, and contribute to making said code base more maintainable. Further study in this direction could prove fruitful.

In our opinion, these results can be worth considering when talking about functional programming languages like Haskell, which do not operate with mutable data at all.

\section{Limitations and Future Plans}\label{sec:limitations}

The usage of a single metric to evaluate the complexity of a Java class is definitely not convincing enough. There are many other metrics, which we plan to consider in future studies, including Lines of Code, Cognitive Complexity~\citep{Campbell2018CognitiveC}, Maintainability Index~\citep{coleman1994}, and Halstead metrics~\citep{halstead1977elements}.

Even though the size of our data seems reasonably big (more than 800 thousand Java classes from 1000 open GitHub repositories), it only represents a tiny fraction of the entire open source Java code landscape. Moreover, there are proprietary Java projects, which we didn't consider in our research. This also makes our conclusions less convincing. In future studies we are planning to analyze proprietary code repositories.

Due to the absence of a formal definition of immutability in existing literature, we gave our own definition. Obviously, this makes the conclusions of our study less convincing, especially for those who have their own definitions of immutability. In future studies, we will try to repeat the analysis with a number of immutability definitions and compare the results.

\section{Conclusion}\label{sec:conclusion}

We analyzed 862,446 Java classes from 1,000 open GitHub repositories and calculated the correlation between their Cyclomatic Complexity (\mtc{CC}) and the number of mutable attributes within them. As a result, we observed that classes with mutable attributes generally have a higher \mtc{CC}, which may indicate lower usability and maintainability of the code. While the average \mtc{CC} of immutable classes was 6.5, the average \mtc{CC} of mutable classes was significantly higher at 19.8. Nevertheless, the extent to which a class is immutable proved to have almost no connection to its \mtc{CC}, with the correlation being -0.02 (0.0003 if static attributes are included). This means that even one violation of the immutability convention resulted in a higher \mtc{CC}, with subsequent violations having little to no impact.

\section{Data Availibility}\label{sec:data_availibility}

All the data and scripts used in this research can be accessed via links in the corresponding citations, which are included whenever such an artifact is first mentioned in the paper. 

\bibliographystyle{plainnat}
\bibliography{main}

\end{document}